\documentclass[prodmode]{acmsmall}

\DeclareMathSizes{10}{8}{6}{4}
\usepackage[ruled]{algorithm2e}
\usepackage{color}
\usepackage{amsmath}
\usepackage{subfigure}
\usepackage{multirow}
\renewcommand{\algorithmcfname}{ALGORITHM}
\SetAlFnt{\small}
\SetAlCapFnt{\small}
\SetAlCapNameFnt{\small}
\SetAlCapHSkip{0pt}
\IncMargin{-\parindent}

\acmVolume{x}
\acmNumber{x}
\acmArticle{xx}
\acmYear{xxxx}
\acmMonth{3}

\begin{document}

\markboth{Y. Shi, M. Larson, and A. Hanjalic}{
Exploiting Social Tags for Cross-Domain Collaborative Filtering}

\title{Exploiting Social Tags for Cross-Domain Collaborative Filtering}
\author{YUE SHI
\affil{Delft University of Technology}
MARTHA LARSON
\affil{Delft University of Technology}
ALAN HANJALIC
\affil{Delft University of Technology}}

\begin{abstract}
One of the most challenging problems in recommender systems based on the
collaborative filtering (CF) concept is data sparseness, i.e., limited user
preference data is available for making recommendations. Cross-domain
collaborative filtering (CDCF) has been studied as an effective mechanism to
alleviate data sparseness of one domain using the knowledge about user
preferences from other domains. A key question to be answered in the context of
CDCF is what common characteristics can be deployed to link different domains
for effective knowledge transfer. In this paper, we assess the usefulness of
user-contributed (social) tags in this respect. We do so by means of the
\textit{Generalized Tag-induced Cross-domain Collaborative Filtering} (GTagCDCF)
approach that we propose in this paper and that we developed based on the
general collective matrix factorization framework. Assessment is done by a
series of experiments, using publicly available CF datasets that represent three
cross-domain cases, i.e., two two-domain cases and one three-domain case. A
comparative analysis on two-domain cases involving GTagCDCF and several
state-of-the-art CDCF approaches indicates the increased benefit of using social
tags as representatives of explicit links between domains
for CDCF as compared to the implicit links deployed by
the existing CDCF methods.
In addition, we show that users from different domains can already benefit from GTagCDCF if they
only share a few common tags. Finally, we use the three-domain case to validate
the robustness of GTagCDCF with respect to the scale of datasets and the varying
number of domains.
\end{abstract}




\acmformat{Shi, Y., Larson, M., and Hanjalic, A.  xxxx.
Exploiting Social Tags for Cross-Domain Collaborative Filtering.}

\begin{bottomstuff}
Portions of this work are based on the paper ``Tags as bridges between domains:
Improving recommendation with tag-induced cross-domain collaborative
filtering"~\cite{Shi2011a}, which appeared in Proceedings of UMAP'11, Girona, Spain, July 2011.

Author's addresses: Y. Shi, M. Larson, A. Hanjalic, Multimedia Information
Retrieval Lab, Department of Intelligent Systems, Faculty of Electrical
Engineering, Mathematics and Computer Science, Delft University of Technology, Mekelweg 4, 2628CD Delft, The
Netherlands; e-mail: \{y.shi, m.a.larson, a.hanjalic\}@tudelft.nl.
\end{bottomstuff}

\maketitle

\section{Introduction}
\label{sec:intro}
Recommender systems have become ubiquitous and play an important role in
relieving internet users from the online information overload. One of the most successful recommendation techniques
is collaborative filtering (CF)~\cite{Adomavicius2005}, which is
based on the assumption that users who have similar interest in the past may still
have similar preference in the future~\cite{Resnick1994}. One of the most challenging problems in
CF is data
sparseness~\cite{Adomavicius2005,Cacheda2011,Herlocker2004}. This challenge 
arises because most of the users in a recommender system do not rate (or give any
kind of feedback on) many items, resulting in highly sparse user-item relations.
With the aim to address the data sparseness problem, recent research in recommender systems has started to study the potential of \textit{cross-domain collaborative
filtering} (CDCF)~\cite{Li2009,Li2009a,Pan2010,Pan2011,Zhang2010}. The underlying
idea of CDCF is to make use of the information common to different
recommender domains in order to compensate for the sparseness in the target domain and benefit the quality of recommendations there. 

Although the idea of CDCF is intuitively sound, the possibilities for its realization are not always straightforward. This is because the information shared by different domains, if available, is not always easily extractable. Different recommender domains are often mutually exclusive, e.g. by focusing on different types of products, like movies, music or books, and by attracting different users. However, even in such cases, CDCF could be possible. For instance, assuming that the domains contain items that could be brought in relation to each other (e.g. like books and movies via genres) and if the ratings follow the same scale, then the rating patterns for items in one domain could propagate to the target domain~\cite{Li2009a}. 

\begin{figure}[t]
\centerline{
	\includegraphics[width=10cm]{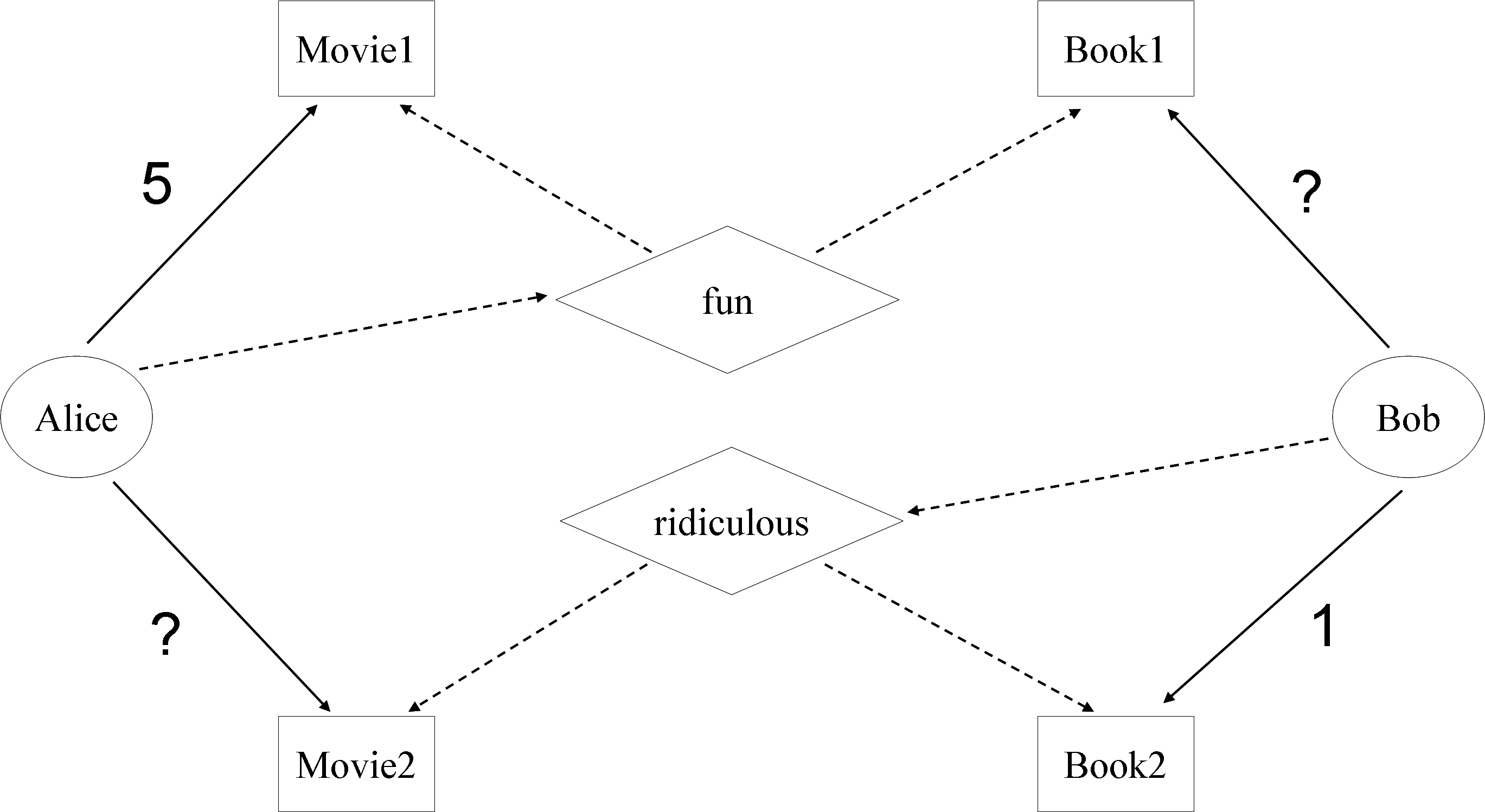}
}
\caption{A toy example of two recommender domains with common tags}
\label{fig:toy}
\end{figure}

In this paper we explore another possible source of common information for CDCF, namely the metadata that are used to describe items in different domains. In particular, we focus on user-generated tags, further referred to as \textit{social tags} and motivate this choice by the following example illustrated in Fig.~\ref{fig:toy}.
Supposing that Alice and Bob are users in \textit{movies} and \textit{books} domains, respectively, we wish to predict Alice's rating on Movie2 and Bob's
rating on Book1, or in other words, whether the movie and book recommender systems should recommend Movie2 and Book1 to Alice and Bob, respectively. Note that in this example we use a solid line to denote
a rating (e.g., scaled from 1 to 5), and a dashed line to denote a relationship
between a user and a tag (user inserted the tag) or a relationship between an item and a tag (tag assigned to item). While two domains may be sparse and it may be difficult to infer recommendations within an individual domain, missing user's preferences could be inferred based on the tags ``fun'' and ``ridiculous'' that are used in both domains. Alice rated 
Movie1 with 5 and also tagged it with ``fun''. We may generalize this by saying that an item tagged by ``fun'' is likely to be
favored by users. Based on this generalization, we can infer that in the book domain,
Bob might also be in favor of Book1, which some other user also tagged with ``fun''. According to the same reasoning, we can infer that Movie2 may not be a good recommendation for Alice. 

While the example in Fig.~\ref{fig:toy} only gives a simplified picture of a
real recommendation scenario, it shows the potential of tags to facilitate CDCF.
Moreover, compared to the approaches exploiting the common rating patterns,
common tags enable CDCF also between domains whose items are not necessarily
related, like books and movies are via genres. Finally, the value of social tags
for CDCF also increases with their growing availability in the recommendation
context. Tagging has become a ubiquitous function in most modern recommender
systems~\cite{Robu2009,Song2011}, allowing users to annotate an item with an
arbitrary textual expression in parallel to adding the ratings.

In order to explore the potential of social tags to its full extent, we
developed the \textit{Generalized Tag-induced Cross-Domain Collaborative
Filtering} (GTagCDCF) approach that we present in this paper. GTagCDCF builds on
the widely adopted concept of collective matrix factorization
(CMF)~\cite{Singh2008}. We choose CMF as the basis of our approach not only
because it has proven to be one of the most effective ways of building
recommender systems based on the CF concept, but also because CMF can elegantly
and effectively be expanded to a multi-domain case, as we will show in
Section~\ref{sec:alg}. GTagCDCF is evaluated in a series of experiments, using
publicly available CF datasets that represent three cross-domain cases, i.e.,
two two-domain cases and one three-domain case. A
comparative analysis on two-domain cases involving GTagCDCF and several
state-of-the-art CDCF approaches indicates the increased benefit of using social
tags as representatives of explicit links between domains
for CDCF as compared to the implicit links deployed by
the existing CDCF methods. In addition, we
show that users from different domains can already benefit from GTagCDCF if they only
share a few common tags. Finally, we use the three-domain case to validate the
robustness of GTagCDCF with respect to the scale of datasets and the varying
number of domains.

The remainder of the paper is structured as follows. In Section~\ref{sec:rw}, we
summarize related work and position our approach with respect to it. We
formalize the research problem addressed in the paper and introduce the related terminology and notation in Section~\ref{sec:prob}. This is followed by
a detailed presentation of GTagCDCF in Section~\ref{sec:alg}. In
Section~\ref{sec:experiment}, we define a number of research questions and present experimental study
on three cross-domain cases that provides answers to these questions. The last section summarizes the main insights gained in this paper about the suitability of social tags for CDCF and sketches the directions for future work.

\section{Related Work}
\label{sec:rw}
This section briefly summarizes the existing related research in CF,
CDCF and tag-aware recommendation, in order to position
the approach we propose in this paper.

\subsection{Collaborative Filtering}
CF approaches are usually categorized as either memory-based or
model-based~\cite{Adomavicius2005,Breese1998,Herlocker2004}. A
recent overview of CF approaches can be found in~\cite{Ekstrand2011}. Depending
on whether the recommendation for a user is aggregated from other users
with similar interests, or from items that are similar to those for which the
user already expressed a preference, memory-based CF approaches can be further
categorized as either user-based collaborative filtering (UBCF)~\cite{Herlocker1999,Resnick1994} or item-based collaborative filtering~\cite{Deshpande2004,Linden2003,Sarwar2001}. Although
various modifications have been made for memory-based approaches, e.g.,
similarity fusion~\cite{Wang2006a} and external
knowledge integration~\cite{Umyarov2011}, the key drawback remains in the
expensive computation for similarities among all users or items, which could hardly scale with the extremely large numbers of users and items in real-world recommender systems. Compared to memory-based approaches,
model-based approaches, which do not rely on computing similarities among users
or items, first use a training set of user-item preference data to learn a
prediction model and then apply that model to generate recommendations.
Conventional model-based CF approaches include Gaussian mixture
model~\cite{Kleinberg2004,Si2003} and latent semantic model~\cite{Hofmann2004}.
Recently, matrix factorization (MF) techniques have attracted much research
attention, due to the advantages of scalability and accuracy, especially for
large-scale data, as exemplified by the Netflix
contest~\cite{Koren2009a,Paterek2007,Salakhutdinov2008}. Generally, MF
techniques exploit the observed user preference data of a recommender system to
learn both latent user features and latent item features, which are further used
to predict unknown user preference for items. Probabilistic matrix factorization
(PMF)~\cite{Salakhutdinov2008} illustrates the rating factorization from a
probabilistic point of view. In
addition, researchers have also proposed to model MF with multiple matrices from
probabilistic point of view by extending the PMF framework with additional
priors~\cite{Adams2010,Porteous2010}.
We note that our work is closely related to the work of MF with multiple
matrices. To the best of our knowledge, under the context of CF, collective
matrix factorization (CMF)~\cite{Singh2008} is the first work that proposed to
factorize multiple matrices related to either users or items in addition to
user-item rating matrix. In the context of text classification, simultaneous factorization of
multiple matrices has already established itself as a notable
approach~\cite{Zhu2007}. The GTagCDCF approach proposed in this paper adopts the concept
of CMF and expands it for the case of multiple domains as explained in Section~\ref{sec:alg}. 

\subsection{Cross-Domain Collaborative Filtering}
Probably the earliest work on CDCF was done by
Berkovsky et al.~\cite{Berkovsky2007}, who deployed several mediation
approaches for importing and aggregating user rating vectors from different
domains. Recently, research on CDCF has been influenced by and
benefited from the progress of transfer learning~\cite{Pan2010a}, a machine
learning paradigm for sharing knowledge among different domains. Coordinate
system transfer~\cite{Pan2010} is proposed to first learn latent features of
users and items from an auxiliary domain (which has relatively more user
preference data) and then adapt them to a target domain (which has relatively
less user preference data) in order to regularize the learning of latent
features for the users and the items in the target domain. Further extension has
also been proposed to exploit implicit user feedback, rather than explicit user
ratings, to constitute the auxiliary domain~\cite{Pan2011}.
One of the recent contributions exploited tensor factorization
models for CDCF~\cite{Hu2013}, in which latent features of domains are learned
in addition to those of users and items. Another recent contribution exploited
the explicit links derived from users' online social network for CDCF~\cite{Shapira2013}. However,
these approaches require that either users or items are shared between the domains, which is, as already mentioned in Section~\ref{sec:intro}, a condition not commonly encountered in practical applications.

Existing approaches addressing disjoint domains (i.e. no common users nor items) include codebook transfer (CBT)~\cite{Li2009} and rating-matrix
generative model (RMGM)~\cite{Li2009a}. These approaches aim at transferring the knowledge between domains by learning an implicit cluster-level rating pattern from an auxiliary domain that is then adopted in the target domain. Similarly, multi-domain CF~\cite{Zhang2010}~\cite{Tang2011} is proposed to
extend PMF in multiple domains by means
of learning an implicit correlation matrix, which links different
domains for knowledge transfer. 
Our proposed GTagCDCF also falls in the last category as it also focuses on the
disjoint recommendation domains where CDCF challenge is largest. However, as
opposed to the concepts mentioned above, we rely on social tags as information
source that links the domains for knowledge transfer. The existing approaches
can be said to rely on hidden rating patterns that can propagate from one domain
into another one only if the items in both domains can be brought in relation to
each other, like for instance when genre is used to characterize both movies and
books. In these cases, it is often attempted to also benefit from hidden social
relations between different user groups that are active in different
domains~\cite{Coyle2008}.
Also, in order to propagate rating patterns, rating should be done in a similar fashion in each domain, e.g. using the same
rating scale. By relying on social tags, CDCF can be applied in a broader range
because no hidden relationship between users or items between different domains
are required. Also, it is not required that user express their preferences in
different domains in the same way. Furthermore, as shown by the experimental
study later in this paper, for successful CDCF, only some minimum number of tags
need to be shared between the domains.

The proposed approach is a generalization of the \textit{tag-induced
cross-domain collaborative filtering} (TagCDCF) that we proposed
before~\cite{Shi2011a}. Specifically, GTagCDCF addresses two drawbacks of
TagCDCF. First, TagCDCF links different domains based on cross-domain
similarities, which need to be computed by using a specific similarity function.
However, the choice of the similarity function might be sensitive to the
recommendation performance, and might vary substantially for different use
cases. Making a good choice from a number of similarity functions usually
requires costly empirical investigation. In addition, calculating these
similarities offline is still a computationally expensive task in the case of
large recommender domains. Second, TagCDCF only exploits user-tag relations and
item-tag relations with binary indicators, while the frequency of a user
assigning a tag and the frequency of an item annotated by a tag are not taken
into account. As a result, the potential of knowledge transfer by the common
tags may not be exploited to its full extent. GTagCDCF does not require
computing cross-domain similarities and does explicitly take into account the
frequency of user-tag relations and item-tag relations.

\subsection{Tag-aware Recommendation}
The work presented in this paper is also related to a general category of CF approaches that we refer to as 
\textit{tag-aware recommendation}. There, tags are exploited to improve
recommendation in various ways, e.g., by integrating
tags into traditional user-based CF and item-based
CF~\cite{Liang2010,Sen2009,Tso-Sutter2008}, by incorporating tags into
probabilistic latent semantic model~\cite{Hofmann2004} to unify user-item relations and item-tag
relations~\cite{Wang2006,Wetzker2009}, and by using tag-based user correlations
as a regularization for PMF~\cite{Zhen2009}. More recently, another group
of state-of-the-art approaches has employed tensor factorization
techniques~\cite{Kolda2008} for tag-aware recommendation. Under such approaches,
item recommendations or tag recommendations are learned from the \{user, tag,
item\} triplet/ternary data directly~\cite{Rendle2010a,Symeonidis2010}. Given
the advantages of using tags for improving recommender systems, our work in this
paper goes a step further and exploits the potential of tags to
introduce mutual benefits between different recommender domains.

\section{Problem Statement, Terminology and Notations}
\label{sec:prob}
The research problem we study in this paper can be stated as: \emph{How to
effectively exploit the common tags between different recommender domains for
improving the quality of recommendations in each domain.} We provide an answer to this question by means of our proposed GTagCDCF approach that will be introduced in Section~\ref{sec:alg}. Before introducing the approach itself, however, we first define the terms and notations used further in the paper and reformulate the above research question accordingly. 

A user-item matrix in $k$th ($k=1,2,\ldots, K$) domain is denoted as $\bf
R^{(k)}$ and is formed using the feedback data collected from $M_k$ users on $N_k$ items, where feedback data can be either implicit or explicit (e.g. ratings). We use $R_{ij}^{(k)}$ to denote user $i$'s preference score derived from feedback data on item $j$ in the $k$th domain. A
user-tag matrix in the $k$th domain is denoted as $\bf F^{(U_k)}$, where
$F_{il}^{(U_k)}$ indicates the frequency of user $i$ using tag $l$. Similarly, an item-tag matrix in $k$th domain is denoted
as $\bf F^{(V_k)}$, where $F_{jl}^{(V_k)}$ indicates the frequency of item
$j$ annotated by tag $l$. Note that without loss of generality the non-zero
entries of $\bf R^{(k)}$, $\bf F^{(U_k)}$ and $\bf F^{(V_k)}$ are, respectively,
normalized to be within the range [0,1] by dividing over the maximal value in each matrix.
$\bf U^{(k)}$ denotes a $d\times M_k$ matrix, whose $i$th column, i.e., $U_i^{(k)}$,
represents a $d$-dimensional latent feature vector of user $i$ in the domain $k$. Similarly, $\bf
V^{(k)}$ denotes a $d\times N_k$ matrix, whose $j$th column i.e., $V_j^{(k)}$,
represents a $d$-dimensional latent feature vector of user $j$ in the domain $k$. Supposing
there are $L$ tags that are common to (i.e., co-exist in) all the $K$ domains,
$\bf T$ denotes a $d\times L$ matrix, whose $l$th column, i.e., $T_l$ represents a
$d$-dimensional latent feature vector of tag $l$. Our objective is to estimate the
unknown latent features of users and items, i.e., $\bf U^{(k)}$ and $\bf
V^{(k)}$ ($k$=1,2\ldots,$K$), which can be then used to predict recommendations
in each domain.
The set of latent
features of common tags, i.e., $\bf T$, takes role of the bridge that
connects $K$ domains. As a result, the latent features of users in one
domain would influence those in other domains via the common tags, and so do
the latent features of items. The strength of the mutual influence is reflected
by the frequency of user-tag relations, i.e., $\bf F^{(U_k)}$ and the frequency
of item-tag relations, i.e., $\bf F^{(V_k)}$. In the following, we will present
the details of learning latent features of users and items.
In addition, we adopt the convention of denoting the number of
non-zero entries in a matrix \textbf{A} as $|$\textbf{A}$|$. We use
$I^A$ as an indicator function, which gives $I_{ij}^A=1$ if $A_{ij}>0$, 0
otherwise. $\|{\bf A}\|_{Fro}$ denotes the Frobenius norm of matrix \textbf{A}.

\section{Descrption of the GTagCDCF approach}
\label{sec:alg}
We first introduce the model and the learning algorithm underlying our GTagCDCF approach, from which we demonstrate how tags bridge different domains. Then, we
present complexity analysis of GTagCDCF and a compact formulation of this
model that may be of use for practitioners. 

\subsection{Model and Learning}
Due to our choice to adopt the widely used concept of CMF to incorporate social tags in CDCF, we proceed by simultaneously factorizing the
user-item rating matrices of $K$ domains, the user-tag matrices of $K$ domains,
and the item-tag matrices of $K$ domains. As a result, the objective function of
GTagCDCF can be formulated as below:
\begin{align}
\label{eqn:obj}
&G({\bf U^{(1)},U^{(2)}\ldots,U^{(K)},V^{(1)},V^{(2)}\ldots,V^{(K)}, T})\\
\nonumber = &\frac{1}{2}\sum_{k=1}^K \sum_{i=1}^{M_k} \sum_{j=1}^{N_k}
\left
[I_{ij}^{R^{(k)}} \left
(R_{ij}^{(k)}-g(U_i^{(k)T}V_j^{(k)}) \right )^2\right ]
 +\frac{\alpha}{2} \sum_{k=1}^K \sum_{i=1}^{M_k} \sum_{l=1}^{L}
\left
[I_{il}^{F^{(U_k)}}\left
(F_{il}^{(U_k)}-g(U_i^{(k)T}T_l) \right )^2 \right ]
\\ \nonumber
& +\frac{\beta}{2} \sum_{k=1}^K \sum_{j=1}^{N_k} \sum_{l=1}^{L}\left
[I_{jl}^{F^{(V_k)}} \left (F_{jl}^{(k)}-g(V_j^{(k)T}T_l) \right )^2\right
]
+ \frac{\lambda}{2}\left [\sum_{k=1}^K (\|{\bf U^{(k)}}\|_{Fro}^2+\|{\bf
V^{(k)}}\|_{Fro}^2)+\|{\bf T}\|_{Fro}^2 \right ]
\end{align}
where 
$\lambda$ is a regularization parameter that penalizes the magnitude of latent features in order to alleviate over-fitting. $\alpha$ and $\beta$ are regarded as tradeoff parameters, which control the relative influence from the user-tag matrices in $K$ domains and the item-tag matrices in $K$ domains, respectively. We emphasize that in the objective function the second term reflects that the latent features of users from the $K$ domains are associated through latent features of shared tags, and the third term reflects that the latent features of items from the $K$ domains are also associated through the latent features of shared tags. Therefore, the proposed GTagCDCF bridges different domains by exploiting the common tags that have relationships with both users and items from different domains. This aspect of influencing latent features of users and items across $K$ domains via tags also presents the main difference
of GTagCDCF from the original CMF. Note that in the case of $K=1$, i.e., a single domain case, GTagCDCF returns to the exact form of CMF.
 
The objective function in Eq.~(\ref{eqn:obj}) is not jointly convex to all the
variables of latent features, i.e., $\bf U^{(k)}$, $\bf V^{(k)}$, and $\bf
T$. We choose to apply gradient descent with respect to one of these variables
alternatively and keep all the other variables fixed, in order to attain a local
minimum solution for the objective function. Specifically, the gradients with
respect to each variable can be computed as below:
\begin{align}
\label{eqn:gdu}
&\frac{\partial G}{\partial U_i^{(k)}}=\sum_{j=1}^{N_k}\left [
I_{ij}^{R^{(k)}}\left ( g(U_i^{(k)T}V_j^{(k)})-R_{ij}^{(k)} \right
)g'(U_i^{(k)T}V_j^{(k)})V_j^{(k)} \right ] \\ \nonumber
& \quad + \alpha\sum_{l=1}^{L}\left [
I_{il}^{F^{(U_k)}}\left ( g(U_i^{(k)T}T_l)-F_{il}^{(U_k)} \right
)g'(U_i^{(k)T}T_l)T_l \right ]
+ \lambda U_i^{(k)} \quad k=1,2\ldots,K\\
\label{eqn:gdv}
 &\frac{\partial G}{\partial V_j^{(k)}}=\sum_{i=1}^{M_k}\left [
 I_{ij}^{R^{(k)}}\left ( g(U_i^{(k)T}V_j^{(k)})-R_{ij}^{(k)} \right
 )g'(U_i^{(k)T}V_j^{(k)})U_i^{(k)} \right ] \\ \nonumber
& \quad + \beta \sum_{l=1}^{L}\left [
 I_{jl}^{F^{(V_k)}}\left ( g(V_j^{(k)T}T_l)-F_{jl}^{(V_k)} \right
 )g'(V_j^{(k)T}T_l)T_l \right ]
+ \lambda V_j^{(k)} \quad k=1,2\ldots,K\\
\label{eqn:gdt}
 &\frac{\partial G}{\partial T_l}=
\alpha\sum_{k=1}^{K} \sum_{i=1}^{M_k}\left [ I_{il}^{F^{(U_k)}}\left (
g(U_i^{(k)T}T_l)-F_{il}^{(U_k)} \right )g'(U_i^{(k)T}T_l)U_i^{(k)}
\right ] \\ \nonumber
& \quad + \beta \sum_{k=1}^{K} \sum_{j=1}^{N_k}\left [ I_{jl}^{F^{(V_k)}}\left
( g(V_j^{(k)T}T_l)-F_{jl}^{(V_k)} \right )g'(V_j^{(k)T}T_l)V_j^{(k)}
\right ] + \lambda T_l
\end{align}

For the consideration of better readability, we summarize the learning algorithm
for GTagCDCF in Algorithm~\ref{alg:one}.

\subsection{Complexity Analysis}
\label{sec:complex}
The complexity of the objective function of GTagCDCF in Eq.~(\ref{eqn:obj}) is
$O(d$$\sum_k(|\bf{R^{(k)}}|$+$|\bf{F^{(U_k)}}|$+$|\bf{F^{(V_k)}}|$+$M_k$+$N_k)$+$dL)$.
The complexity of computing each gradient in
Eq.~(\ref{eqn:gdu}$\sim$\ref{eqn:gdt}) is
$O(d$$\sum_k(|\bf{R^{(k)}}|$+$|\bf{F^{(U_k)}}|$+$M_k))$, $O(d$$\sum_k(|\bf{R^{(k)}}|$+$|\bf{F^{(V_k)}}|$+$N_k))$, and
$O(d$$\sum_k(|\bf{F^{(U_k)}}|$+$|\bf{F^{(V_k)}}|)$+$dL)$, respectively. 
 Considering that in reality we usually
have conditions as $|\bf{R_{(k)}}|$$>>M_k, N_k$, $|\bf{F^{(U_k)}}|$$>>M_k,L$,
and $|\bf{F^{(V_k)}}|$$>>N_k,L$, the total complexity of GTagCDCF is in the
order of $O($$\sum_k(|\bf{R^{(k)}}|$+$|\bf{F^{(U_k)}}|$+$|\bf{F^{(V_k)}}|))$,
which is linear in the total number of the known preference scores in
$\bf{R^{(k)}}$, the known user-tag relationships in $\bf{F^{(U_k)}}$, and the
known item-tag relationships in $\bf{F^{(V_k)}}$ from all the domains. This
analysis indicates that GTagCDCF is appropriate for large scale use cases.

\begin{algorithm}[t]
\SetAlgoNoLine
\KwIn{Normalized user-item preference matrix $\bf R^{(k)}$ ($k$=1,2\ldots,$K$),
		normalized user-tag frequency matrix $\bf F^{(U_k)}$ ($k$=1,2\ldots,$K$),
		normalized item-tag frequency matrix $\bf F^{(V_k)}$ ($k$=1,2\ldots,$K$),
tradeoff parameters $\alpha$, $\beta$,
regularization parameter $\lambda$,
stop condition $\epsilon$.}
\KwOut{Predicted user-item preference matrix $\bf \hat{R}^{(k)}$
($k$=1,2\ldots,$K$).} Initialize $\bf U^{(k)[0]}$, $\bf V^{(k)[0]}$
($k$=1,2\ldots,$K$) and $\bf T^{[0]}$ with random values\; 
$t=0$\; 
Compute $G^{[t]}$ as in Eq.~(\ref{eqn:obj})\;
\Repeat{$f\leq \epsilon$}{
\For{$k=1,2,\ldots,K$}{
$\eta=1$\;
Compute $\frac{\partial G}{\partial U^{(k)[t]}}$, $\frac{\partial G}{\partial
V^{(k)[t]}}$ and $\frac{\partial G}{\partial T^{[t]}}$ as in
Eq.~(\ref{eqn:gdu}$\sim$\ref{eqn:gdt}) \; 
\Repeat{$G({\bf U^{(k)[t]}}-\eta \frac{\partial G}{\partial
U^{(k)[t]}},{\bf V^{(k)[t]}}-\eta \frac{\partial G}{\partial
V^{(k)[t]}},{\bf T^{[t]}}-\eta \frac{\partial G}{\partial T^{[t]}})< G^{[t]}$}{
$\eta=\eta/2$;   // maximize learning step size\\
} 
${\bf U^{(k)[t+1]}}={\bf U^{(k)[t]}}-\eta \frac{\partial G}{\partial
U^{(k)[t]}}$\;
${\bf V^{(k)[t+1]}}={\bf V^{(k)[t]}}-\eta \frac{\partial G}{\partial
V^{(k)[t]}}$\; 
${\bf T^{[t+1]}}={\bf T^{[t]}}-\eta \frac{\partial G}{\partial
T^{[t]}}$\;}
Compute $G^{[t+1]}$ as in Eq.~(\ref{eqn:obj})\;
$f=1-G^{[t+1]}/G^{[t]}$\;
$t=t+1$\;
}
$\bf \hat{R}^{(k)}=U^{(k)[t]T}V^{(k)[t]}$ ($k$=1,2\ldots,$K$)\;
\caption{GTagCDCF}
\label{alg:one}
\end{algorithm}

\subsection{A Compact Formulation}
It is worth mentioning that we can present the objective function of GTagCDCF in
a compact form, which is mathematically equivalent to Eq.~(\ref{eqn:obj}). The
compact formulation is expressed below:
\begin{align}
\label{eqn:compact}
G({\bf X})
= &\frac{1}{2} \sum_{i=1}^{M_0} \sum_{j=1}^{N_0}
\left
[I_{ij}^{\mathbb{R}} \left
(\mathbb{R}_{ij}-g(\mathbb{U}_i^{T}\mathbb{V}_j) \right )^2\right ]
 +\frac{\alpha}{2} \sum_{i=1}^{M_0} \sum_{l=1}^{L}
\left
[I_{il}^{\mathbb{F^{(U)}}}\left
(\mathbb{F}_{il}^{(\mathbb{U})}-g(\mathbb{U}_i^{T}T_l) \right )^2
\right ] \\ \nonumber
& +\frac{\beta}{2} \sum_{j=1}^{N_0} \sum_{l=1}^{L}\left
[I_{jl}^{\mathbb{F}^{(\mathbb{V})}} \left
(\mathbb{F}_{jl}^{(\mathbb{V})}-g(\mathbb{V}_j^{T}T_l) \right )^2\right
] + \frac{\lambda}{2}\left [ (\|{\bf \mathbb{U}} \|_{Fro}^2+\|{\bf
\mathbb{V}}\|_{Fro}^2)+\|{\bf T}\|_{Fro}^2 \right ]
\end{align}
where $M_0=M_1+M_2+\ldots+M_K$, and $N_0=N_1+N_2+\ldots+N_K$. $\mathbb{R}$ is a
block diagonal matrix that contains all the preference data from $K$ domains. $\mathbb{F}^{(\mathbb{U})}$ is a
stacked matrix that contains the user-tag relations from all the users of $K$
domains and $\mathbb{F}^{(\mathbb{V})}$ is a stacked matrix that contains
the item-tag relations from all the items of $K$ domains. Specifically, these
variables are shown below:
\begin{align}
\mathbb{R}=\begin{bmatrix}
{\bf R^{(1)}} &  &  &\\ 
 & {\bf R^{(2)}} &  &\\ 
&  & \ldots &\\
 &  &  & {\bf R^{(K)}}
\end{bmatrix},
\quad \mathbb{F}^{(\mathbb{U})}=\begin{bmatrix}
{\bf F^{(U_1)}}\\ 
{\bf F^{(U_2)}}\\ 
\ldots\\ 
{\bf F^{(U_K)}}
\end{bmatrix},
\quad \mathbb{F}^{(\mathbb{V})}=\begin{bmatrix}
{\bf F^{(V_1)}}\\ 
{\bf F^{(V_2)}}\\ 
\ldots\\ 
{\bf F^{(V_K)}}
\end{bmatrix}
\end{align}
Correspondingly, $\mathbb{U}$ and $\mathbb{V}$ are stacked matrices that
contain the latent features of users and items, respectively, from $K$ domains,
as shown below:
\begin{align}
\mathbb{U}=\begin{bmatrix}
{\bf U^{(1)}}\\ 
{\bf U^{(2)}}\\ 
\ldots\\ 
{\bf U^{(K)}}
\end{bmatrix},
\quad \mathbb{V}=\begin{bmatrix}
{\bf V^{(1)}}\\ 
{\bf V^{(2)}}\\ 
\ldots\\ 
{\bf V^{(K)}}
\end{bmatrix}
\end{align}

We consider that the compact formulation of GTagCDCF could help
practitioners to implement the algorithm more easily.

\section{EXPERIMENTAL EVALUATION}
\label{sec:experiment}
In this section, we report on the experiments we conducted to evaluate the
proposed GTagCDCF approach, through which we assess the utility of social tags as effective bridges between domains facilitating CDCF. We start by giving a detailed description of the datasets that are used in our experiments and explain the experimental setup. This is followed by an analysis of the impact of trade-off parameters $\alpha$ and $\beta$ influencing the relative contributions from the cross-domain user-tag relations
and the cross-domain item-tag relations on the performance of GTagCDCF. We deploy this analysis to optimize the parameter setting of the GTagCDCF algorithm. The rest of the experiments are then performed on the optimized algorithm. We organize the presentation of the results of these experiments to answer the following research questions:
\begin{enumerate}
  \item Is the proposed GTagCDCF effective in improving the recommendation performance in different domains?
  \item How does GTagCDCF perform compared to representative state-of-the-art CDCF
  approaches?
   \item How does GTagCDCF effectiveness depend on the number of tags shared between domains? 
   \item What is the influence of user's individual tagging behavior on the benefit drawn from CDCF?
 \item How robust is GTagCDCF regarding the scale of datasets and the number of involved domains?
\end{enumerate}

\subsection{Datasets}
\label{sec:data}
Our experiments are organized in two parts: one for the
two-domain case and the other for the three-domain case. For the two-domain
case, the experiments are conducted on two pairs of publicly available datasets. The first pair of datasets consists of
one subset from MovieLens 10 million
dataset\footnote{http://www.grouplens.org/node/73}~\cite{Herlocker1999} and one
subset from LibraryThing
dataset\footnote{http://dmirlab.tudelft.nl/users/maarten-clements}~\cite{Clements2010a}.
The original MovieLens dataset contains 10 million ratings from 71576 users and
10681 movies, and in the subset we select the first 5000 users and 5000 movies
according to the identifiers in the original dataset. The original LibraryThing
dataset contains ca. 750 thousand ratings from 7279 users and 37232 books, and
in the subset we also select the first 5000 users and 5000 books. This subset
selection was necessary, since some baselines are too computationally expensive
to tackle larger datasets. Note that our choice of the subset selection
procedure rather than random selection also ensures future experimental
reproducibility. In the following, the two used subsets are denoted as ML1 (from
MovieLens) and LT (from LibraryThing). Both ML1 and LT have 5-star rating scale
with half star increments, representing a case that two different domains both
have explicit ratings. In addition to ratings, the two domains also have 2277
common tags. Our target of experimenting on this pair of datasets is to
investigate the effectiveness of our proposed GTagCDCF for benefiting different
recommender domains that are both based on explicit ratings. For notation
convenience, we further refer to this pair of datasets as \textbf{P1}.
The second pair of datasets are from recent initiatives on
information heterogeneity and fusion in recommender
systems\footnote{http://ir.ii.uam.es/hetrec2011/datasets.html}~\cite{Cantador2011},
where three datasets are provided with various types of user
preferences and resources from three different domains, i.e., movie, webpage and
music. Two datasets are used in our work. One dataset is a subset of
MovieLens 10 million dataset\footnote{http://www.grouplens.org/node/462}, only
including the users who have used
both ratings and tags. This dataset contains
ca. 850 thousand ratings (with the same scale of ML1 as mentioned before) from
2113 users and 10197 movies. Although additional public information from
IMDB\footnote{http://www.imdb.com} and Rotten Tomatoes\footnote{http://www.rottentomatoes.com} websites is available for this
dataset, in our experiments we only use the information of user-item-rating and
user-item-tag. We denote this dataset as ML2. The other dataset is collected from Last.fm\footnote{http://www.last.fm} online music system, mainly consisting of user-artist listening information, i.e., the
frequency that a user listened to songs from a music artist. 
We denote this dataset as LF. LF represents a dataset from implicit-feedback
recommender systems, where no explicit user preferences are expressed. LF
contains ca. 93 thousand user-artist listening relations, i.e., listening
counts, from 1892 users and 17632 artists. In addition, there are in total 996
tags that are common to the two domains. Compared to \textbf{P1}, our target of
experimenting on this pair of datasets is to investigate the effectiveness of our proposed GTagCDCF for
benefiting different recommender domains that have different types of user
preferences. We further refer to this pair of datasets as \textbf{P2}. The statistics of all the datasets involved in the
two-domain experiments are summarized in Table~\ref{tab:data}.
\begin{table}
\tbl{Statistics of datasets in \textbf{P1} and \textbf{P2}. 
\label{tab:data}
}
{%
\begin{tabular}{|c|c c|c c|}
\hline
 & \multicolumn{2}{c|}{\textbf{P1}}  & \multicolumn{2}{c|}{\textbf{P2}} \\
 & \textbf{ML1} & \textbf{LT} & \textbf{ML2} & \textbf{LF}\\
 \hline
\#users & 5000 & 5000 & 2113 & 1892\\
\hline
\#items & 5000 & 5000 & 10109 & 17632\\
\hline
\#preferences & 584628 & 179419 & 855598 & 92834\\
\hline
sparseness & 97.70\% & 99.30\% & 96\% & 99.70\%\\
\hline
\#common tags & \multicolumn{2}{c|}{2277}  & \multicolumn{2}{c|}{996}  \\
\hline
\#relations between users and common tags & 559 & 157932 & 2551 & 9402\\
\hline
\#relations between items and common tags & 10778 & 97277 & 4016 & 28267\\
\hline
\end{tabular}}
\begin{tabnote}%
\end{tabnote}%
\end{table}%

For the three-domain case, the experiments are conducted by using the entire
MovieLens 10 million dataset (denoted as ML-all, representing the movie
domain,), the entire LibraryThing dataset (denoted as LT-all, representing the
book domain) and the LF dataset (representing the music domain). Apart from the 2277 common tags
between the ML-all dataset and the LT-all dataset as shown in
Table~\ref{tab:data}, there are 843 common tags between the ML-all dataset and
the LF dataset, and 1152 common tags between the LT-all dataset and
the LF dataset. 

Note that in the following evaluation, our focus is on the
two-domain case for the purpose of demonstrating the usefulness of the GTagCDCF
model, while the experiments of the three-domain case only serve to validate the
robustness of GTagCDCF to larger scale datasets and use cases with more than two
domains.

\subsection{Experimental Setup}
\label{sec:setup}
\subsubsection{Experimental Protocol for Two-Domain Case}
The experimental protocol we devised for the two-domain case follows a similar rationale as in the related CDCF work (e.g. ~\cite{Li2009,Li2009a}). In our experiments, we split each of all
the datasets in both \textbf{P1} and \textbf{P2} into three sets containing
different users, i.e., a training set, a validation set and a test set. For each
dataset, the training set contains 60\% randomly selected users and their preferences (ratings or listening counts) to items (movies, books or artists). The validation set contains 20\% randomly selected users and
their preferences on items. The test set contains the remaining 20\% users and
corresponding preferences. The validation set is used to investigate the impact of
different parameters in the proposed GTagCDCF algorithm and also tune parameters
for all the baseline approaches. The test set is used to evaluate the
performance of GTagCDCF and compare it with other baselines. For the purpose of
investigating the performance of GTagCDCF for users with different
rating profiles, we hold out preferences of users in the test set to be
predicted according to different conditions of user profile length (UPL). For example, under the
condition of UPL=5, we use 5 randomly selected preferences for each user in the
test set. The user profiles of length 5 are taken together with the training
set as training data, and we use the remaining preferences of the users in the
test set for evaluation. In our experiments, we evaluate three different conditions of UPL, i.e., UPL=5, 10, 15. For each condition of
UPL, we generate 10 test data folds (each fold consisting of known preferences
and holdout preferences) by random selection, and we report performance as an
average across all 10 folds. In addition, in order to guarantee that each
user in the test set has sufficient holdout preferences to be evaluated, we
filter out users in the test set who have less than 20 preferences. Therefore,
even for the condition of UPL=15, we still have at least 5 items for each user
in the test set for evaluation.

\subsubsection{Experimental Protocol for Three-Domain Case}
As mentioned in Section~\ref{sec:data}, the purpose of the experiments for the
three-domain case is only to validate the robustness of GTagCDCF. For this
reason, we simply evaluate the performance of GTagCDCF by cross-validation
without detailed examination of different user profiles. Specifically, for each
of the three datasets, we split it into 5 disjoint folds (i.e., each fold
contains 20\% preference data of the whole corresponding dataset.), from which
we randomly select one for tuning the parameters in GTagCDCF and in the baseline approaches, and use the other four folds for cross-validation. The reported performance is
averaged across all the four folds of each dataset.

\subsubsection{Evaluation Metric}
To be consistent with the majority of recent related work on
CDCF~\cite{Li2009,Li2009a,Pan2010,Pan2011}, we adopt mean absolute error (MAE)
as the evaluation metric for measuring recommendation performance on rating-based recommender domains, i.e.,
ML1 and LT in \textbf{P1} and ML2 in \textbf{P2}. Specifically, the definition
of MAE is expressed as below:
\begin{align}
\label{eqn:mae}
MAE=\sum_{(i,j)\in T_E}|\hat{R}_{ij}-R_{ij}|/|T_E|
\end{align}
in which $T_E$ denotes the set of user-item pairs whose ratings need to be
predicted, and $|T_E|$ denotes the number of user-item pairs in the set.
$\hat{R}_{ij}$ denotes the predicted rating for user $i$ to item $j$, and
$R_{ij}$ the corresponding ground truth. A lower value MAE indicates a
better recommendation performance.

For measuring recommendation performance on the LF dataset, whose domain
includes only implicit user feedback, i.e., listening counts, we choose to
use mean average precision (MAP)~\cite{Herlocker2004}, an evaluation metric that is widely used to
evaluate the quality of a ranking list. Specifically, MAP is defined as below:
\begin{align}
\label{eqn:map}
MAP=\frac{1}{M_{te}}\sum_{i=1}^{M_{te}}\frac{\sum_{j=1}^{N_i}(rel_i(j)\times
P_i@j)}{\sum_{j=1}^{N_i}rel_i(j)}
\end{align}
where $M_{te}$ is the number of users for evaluation, and $N_i$ denotes the
number of recommended items (i.e., artists in LF dataset) for the user $i$.
$rel_i(j)$ is a binary indicator, which is equal to 1 if the item of rank $j$ is relevant to user $i$, and is
equal to 0 otherwise. $P_i@j$ is the precision of the top $j$ recommended items
for the user $i$, i.e., the ratio of movies in the top $j$ recommendation that are
relevant to the user $i$. Higher values of MAP indicate a better recommendation
performance. Note that in order to determine the relevance of each
artist to a given user in the test set, we set a threshold of listening counts
for each test user proportional to the maximal listening count in her holdout
set of feedbacks (listening counts to artists). The proportion coefficient for
the threshold is set to 0.7 in our work. For example, if a holdout set of
feedbacks for a test user has a maximal value as 1000, i.e., the maximal number
of listening counts for that user to the artists that are to be ranked is 1000,
then we regard the artists who are listened to no less than 700 (i.e., 0.7$\times$
1000) times by the user as relevant ones. Although other strategies can be
considered as well to determine the relevance of artists, the strategy choice does not influence the
result of our comparative study as long as it matches the
assumption that the user would prefer one artist to the other if she listened to
the one more than the other. In addition, we note that our evaluation is only based
on the artists who were listened to by a given test user. The artists that were not
listened to by the user are not taken into account in the recommendation list,
since there is no ground truth of user preference to those artists. Consequently, our
evaluation on LF dataset is conservative, which could
underestimate the power of recommendation approaches, as shown later in
section~\ref{sec:perform}. However, this issue does not influence our
comparative experimental evaluation, where we are primarily interested in the relative strength of different
approaches.

\subsubsection{Parameter Setting}
In the following experiments, we set the dimensionality of latent
features in GTagCDCF to 10. We notice that, just like in a common MF
technique~\cite{Weimer2008}, the performance of GTagCDCF did not substantially
change when further increasing $d$, while more computational cost is required
(cf. section~\ref{sec:complex}). The regularization parameter $\lambda$ of
GTagCDCF is set to 0.01 for the two-domain case of \textbf{P1}, 0.001
for the two-domain case of \textbf{P2}, and 0.01 for the three-domain case. The
choice of $\lambda$ for each case is based on our observation of the performance of GTagCDCF on the
corresponding validation sets/folds. The impact of $d$ and $\lambda$ on
recommendation performance has been widely investigated in related work
particularly on MF~\cite{Koren2009a,Salakhutdinov2008,Weimer2008}. In the
following, we will focus on the impact of two key parameters of GTagCDCF,namely the tradeoff parameters $\alpha$ and
$\beta$.

\subsection{Impact of Tradeoff Parameters}
\label{sec:para}
\begin{figure}[t]
\centerline{
	\subfigure[ML1]{
	\includegraphics[width=6cm]{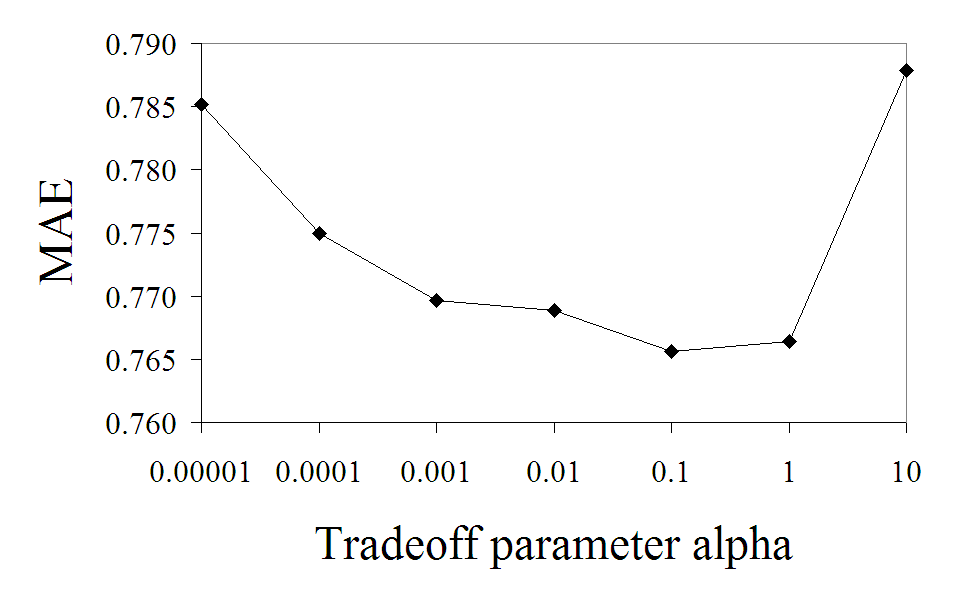}
	}
	\subfigure[LT]{
	\includegraphics[width=6cm]{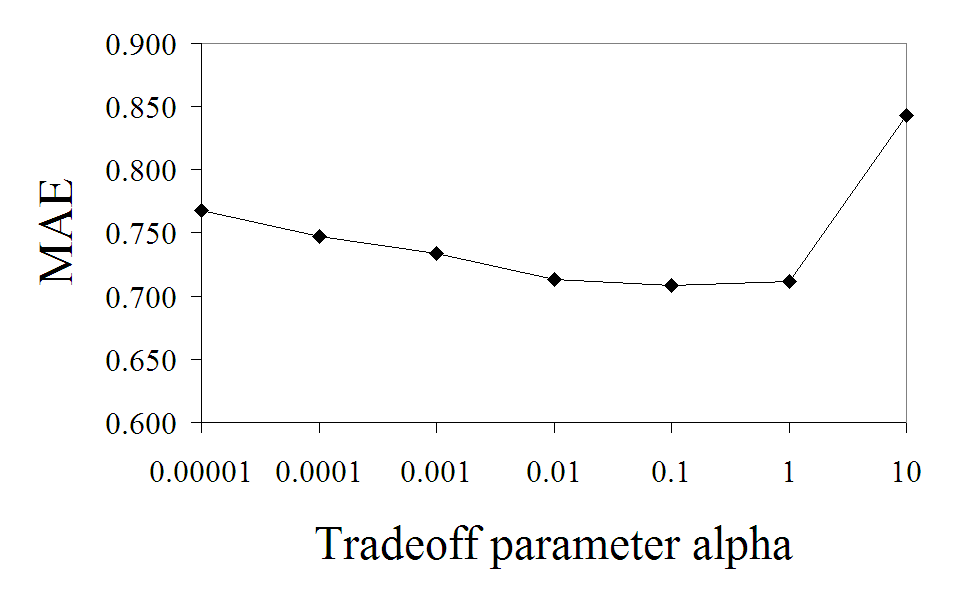}
	}
}
\caption{The impact of tradeoff parameter $\alpha$ on the recommendation
performance of GTagCDCF for the two domains in \textbf{P1}.}
\label{fig:alphap1}
\end{figure}

\begin{figure}[t]
\centerline{
	\subfigure[ML2]{
	\includegraphics[width=6cm]{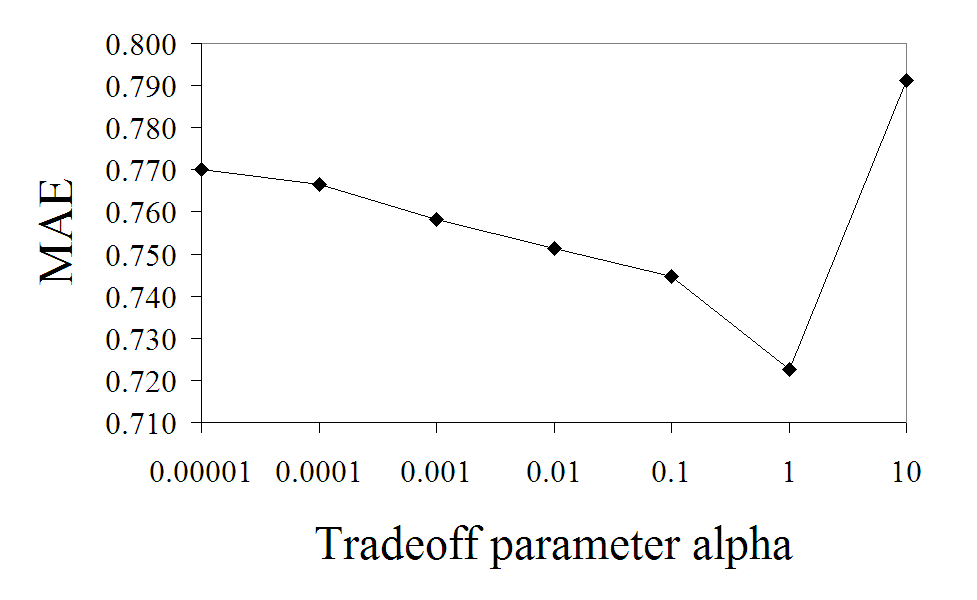}
	}
	\subfigure[LF]{
	\includegraphics[width=6cm]{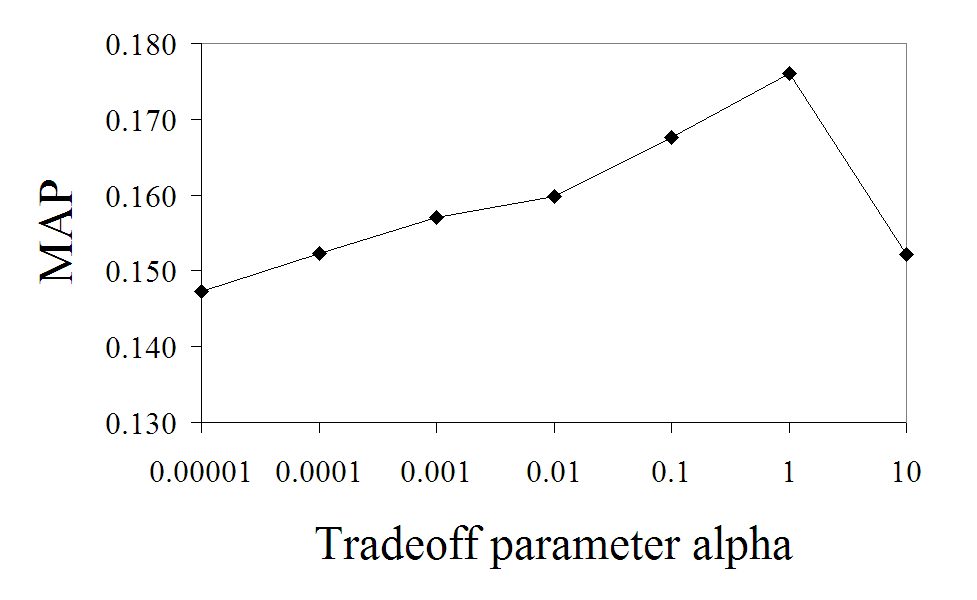}
	}
}
\caption{The impact of tradeoff parameter $\alpha$ on the recommendation
performance of GTagCDCF for the two domains in \textbf{P2}.}
\label{fig:alphap2}
\end{figure}

The tradeoff parameters $\alpha$ and $\beta$ in the proposed GTagCDCF algorithm
influence the relative contributions from the cross-domain user-tag relations
and the cross-domain item-tag relations. By using the validation set together
with the training set in \textbf{P1} and \textbf{P2}, we investigate the
impact of the tradeoff parameters on GTagCDCF by varying their values and
measuring the recommendation performance in terms of MAE on ML1, LT and ML2
datasets and MAP on LF dataset. Note that in the experiments reported in this section, we
set the condition of UPL=5 for the users in the validation set, while measuring
recommendation performance according to the holdout set of ratings or implicit
feedback as mentioned in the previous subsection. We first set $\beta=0$ in
Eq.~(\ref{eqn:obj}) and investigate the impact of $\alpha$, as shown in
Fig.~\ref{fig:alphap1} and Fig.~\ref{fig:alphap2}. 
It can be seen that for the cross-domain case of \textbf{P1}, the optimal value
of $\alpha$ lies around 0.1 in terms of MAE for both ML1 domain and LT domain.
In addition, it can also be seen that there is an optimal value of $\alpha$,
which lies around 1, in terms of MAE in ML2 domain and MAP in LF domain, when
GTagCDCF is used for the case of \textbf{P2}. This observation indicates that, by
exploiting the common tags from different domains, GTagCDCF could be beneficial for
improving latent user features in each domain, thus, resulting in an improved
recommendation performance. 

\begin{figure}[t]
\centerline{
	\subfigure[ML1]{
	\includegraphics[width=6cm]{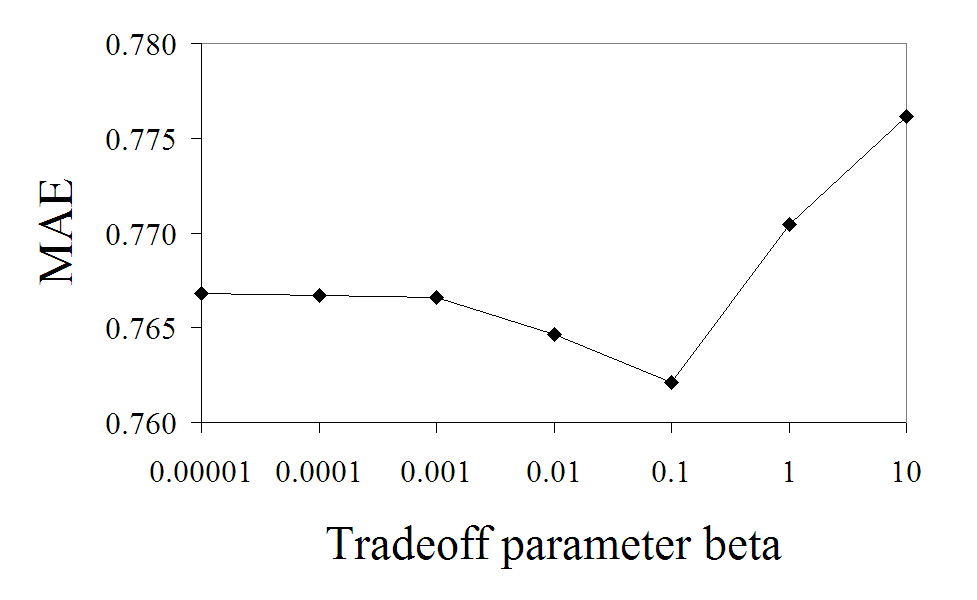}
	}
	\subfigure[LT]{
	\includegraphics[width=6cm]{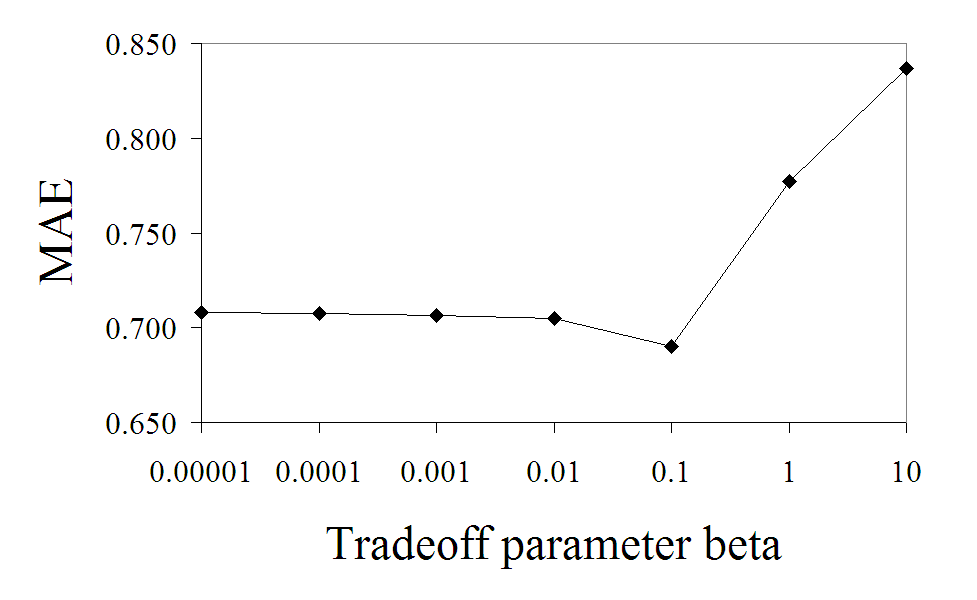}
	}
}
\caption{The impact of tradeoff parameter $\beta$ on the recommendation
performance of GTagCDCF for the two domains in \textbf{P1}.}
\label{fig:betap1}
\end{figure}

\begin{figure}[t]
\centerline{
	\subfigure[ML2]{
	\includegraphics[width=6cm]{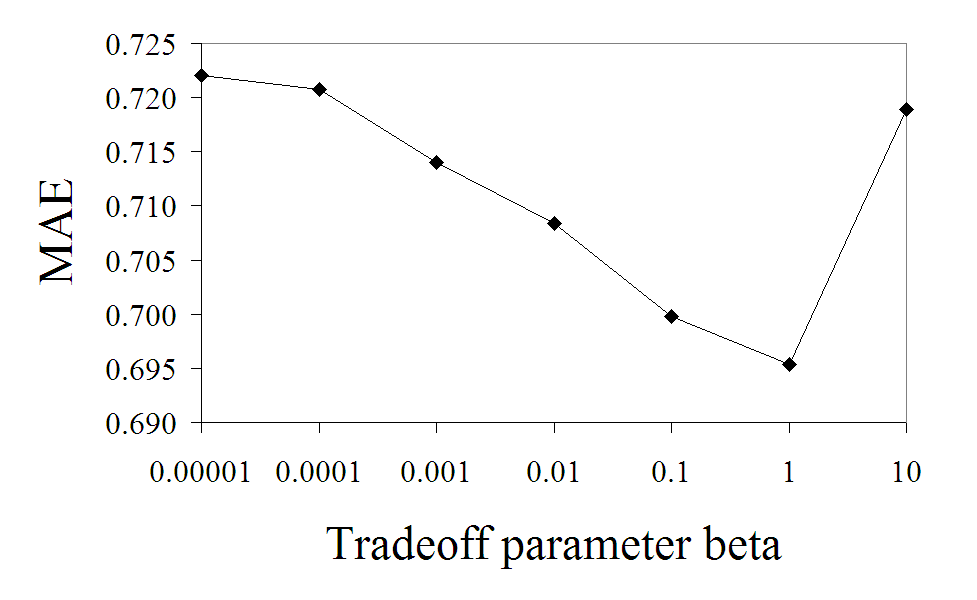}
	}
	\subfigure[LF]{
	\includegraphics[width=6cm]{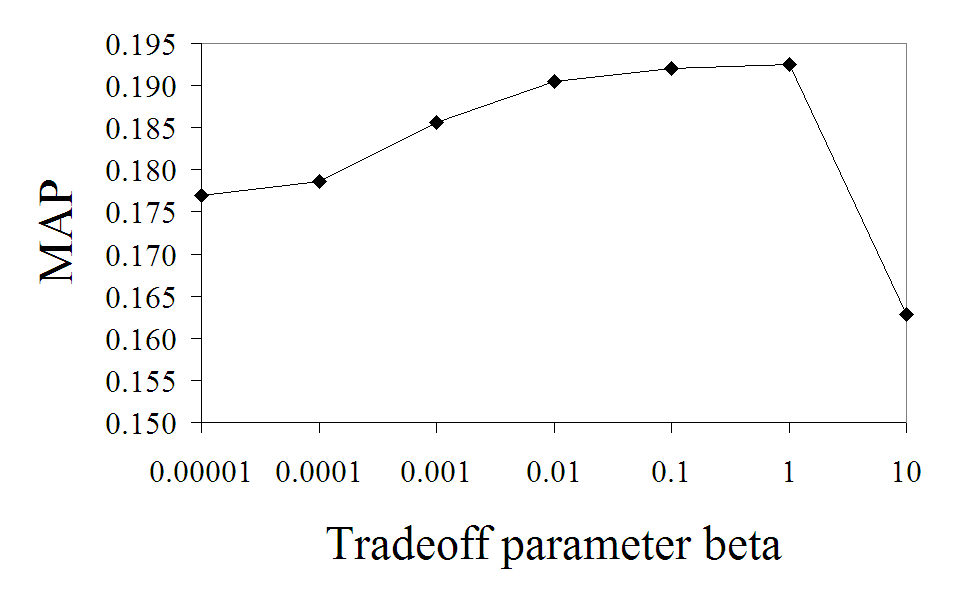}
	}
}
\caption{The impact of tradeoff parameter $\beta$ on the recommendation
performance of GTagCDCF for the two domains in \textbf{P2}.}
\label{fig:betap2}
\end{figure}

By adopting the optimal value of $\alpha=0.1$ for the cross-domain case of
\textbf{P1}, we further investigate the impact of tradeoff parameter $\beta$ in
Eq.~(\ref{eqn:obj}) on the recommendation performance of GTagCDCF. We also
further investigate the impact of tradeoff parameter $\beta$ on GTagCDCF for the
cross-domain case of \textbf{P2}, with $\alpha$ fixed to the optimal value 1.
The results are shown in Fig.~\ref{fig:betap1} and Fig.~\ref{fig:betap2}. As can
be seen in Fig.~\ref{fig:betap1}, the
optimal value of $\beta$ lies around 0.1 in terms of MAE for both domains in
\textbf{P1}. In addition, it can be also seen in Fig.~\ref{fig:betap2} that the
optimal value of $\beta$ lies around 1 in terms of MAE in ML2 domain and MAP
in LF domain. This indicates that GTagCDCF could also be beneficial
for improving latent item features in each domain in addition to latent user
features as shown above, thus, resulting in a further improvement of the recommendation performance in
each domain.


\subsection{Effectiveness}
\begin{figure}[t]
\centerline{
	\subfigure[P1]{
	\includegraphics[width=6cm]{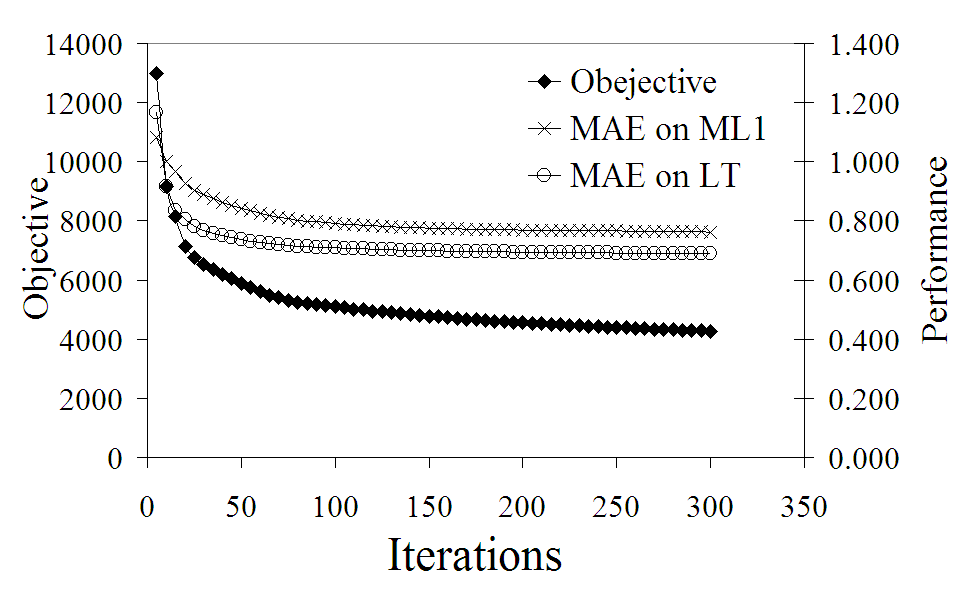}
	}
	\subfigure[P2]{
	\includegraphics[width=6cm]{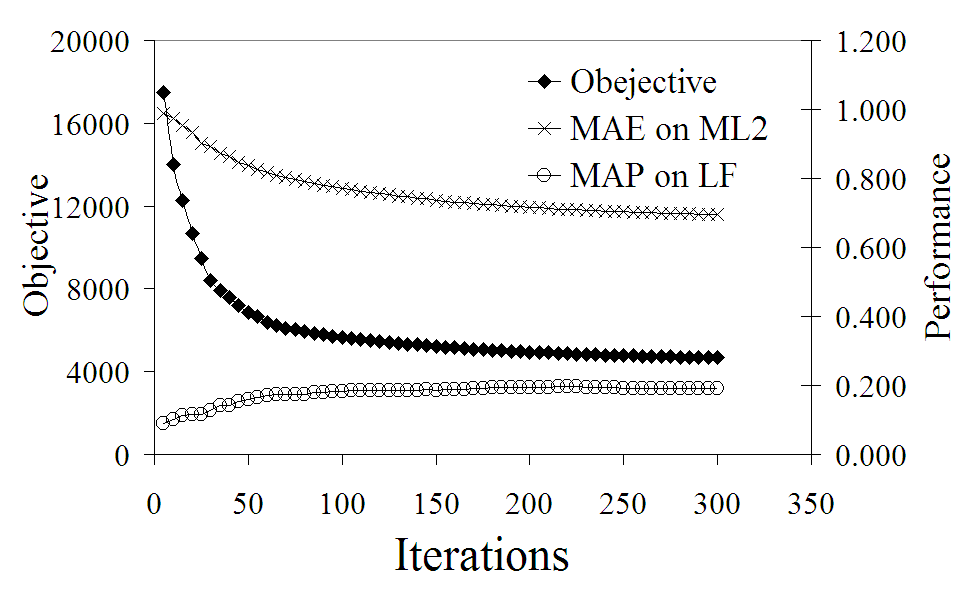}
	}
}
\caption{The effectiveness of GTagCDCF in improving the recommendation
performance (MAE or MAP) in different domains.}
\label{fig:effect}
\end{figure}

In the second set of our experiments, we investigate the effectiveness of
GTagCDCF, i.e., the effect of the minimization of the objective
function in Eq.~(\ref{eqn:obj}) on the improvement of the
recommendation performance in different domains. Note that these
experiments are also based on the validation sets under the condition of UPL=5,
and the tradeoff parameters are adopted with the optimal values for
\textbf{P1}and \textbf{P2}, respectively, as observed from the previous
subsection. To this end, we demonstrate the variation of the output of the
objective function and evaluation
metrics, i.e., MAE for ML1, LT and ML2, and MAP for LF on the validation sets,
simultaneously during the iterations of the optimization process. The results
are shown in Fig.~\ref{fig:effect}. As can be seen, when minimizing the
objective function of GTagCDCF, the recommendation performance in each domain
(for both \textbf{P1} and \textbf{P2}) improves along with algorithm iterations,
approaching convergence after 100 iterations. This observation, supported by the conclusions drawn in previous section regarding the potential of exploiting shared tags to improve both latent item and user features in each domain, allows us to confirm the effectiveness of the proposed approach and provide a positive answer to our first research question.

\subsection{Performance Comparison in the Two-domain Case}
\label{sec:perform}
In this subsection, we compare the performance of the proposed GTagCDCF with a
set of alternative recommendation approaches listed below. As mentioned in
section~\ref{sec:setup}, the performance is reported based on the test set with
10 randomly separated folds. The tradeoff parameters are the optimal ones
determined using the validation set as stated in section~\ref{sec:para}, i.e.,
$\alpha=0.1$, $\beta=0.1$ for ML1 and LT in \textbf{P1}, and $\alpha=1$,
$\beta=1$ for ML2 and LF in \textbf{P2}.

\begin{itemize}
\item \textbf{UBCF}: User-based collaborative filtering~\cite{Herlocker1999} is used as a
representative of memory-based CF approaches. The neighborhood size is tuned to
50 according to our observation of its performance on the validation sets.
\item \textbf{PMF}: Probabilistic matrix factorization~\cite{Salakhutdinov2008}
is a state-of-the-art model-based CF approach. The regularization parameter $\lambda$ is tuned based
on the validation set of each dataset, i.e., 0.01 for ML1 and LT, and 0.001 for ML2 and
LF. Note that both UBCF and PMF are CF approaches for a single-domain use case.
\item \textbf{CBT}: Codebook transfer~\cite{Li2009} represents a
state-of-the-art cross-domain CF approach. For the two domains in \textbf{P1},
we use one domain (e.g., ML1) as the auxiliary domain, which is
used to construct a codebook, and the other domain (e.g., LT) as
the target domain in which the recommendations are generated.
Following the experimental protocol used in~\cite{Li2009}, we select 500 users
and 500 items with most ratings to construct
the auxiliary domain, and set the number of clusters to 50 for both users and
items. Note that CBT is not applicable for the case of \textbf{P2}, since it
requires that two domains share a same rating scale.
\item \textbf{RMGM}: Rating-matrix generative model~\cite{Li2009a} represents
another state-of-the-art cross-domain CF approach. The implementation of RMGM is based
on the publicly available code
package~\footnote{http://sites.google.com/site/libin82cn/} supplied by the
authors. As suggested in~\cite{Li2009a}, we set the number of both the user and
the item clusters to be 20. Similar to CBT, RMGM is not applicable for the case
of \textbf{P2}, since it also requires that two domains share a same rating
scale. Note that in both CBT and RMGM, the related parameters are also tuned
based on the performance measured from the validation sets, as used in GTagCDCF.
\item \textbf{TagCDCF}: Tag-induced cross-domain collaborative filtering is
our previously proposed CDCF approach~\cite{Shi2011a}. The difference between TagCDCF
and GTagCDCF has been discussed in Section~\ref{sec:rw}. Note that all the
parameters involved in TagCDCF are tuned by the same procedures as in GTagCDCF and according to observations from the validation sets. 
\end{itemize}

The results of the comparative analysis are shown in Table~\ref{tab:performp1}
and Table~\ref{tab:performp2} and allow us to make several observations regarding
the proposed algorithm GTagCDCF.

First, as the basic quality check, Table~\ref{tab:performp1}
and Table~\ref{tab:performp2} show that GTagCDCF significantly outperforms representative single-domain
CF approaches, i.e., UBCF and PMF. Note that the significance of improvement is
measured according to Wilcoxon signed rank significance test with p$<$0.005. For
the two domains in \textbf{P1}, GTagCDCF improves over UBCF by 7$\sim$8\% on
ML1 and 14$\sim$19\% on LT, and over PMF by up to 8.5\% on ML1 and up to
10\% on LT. Similar improvement can also be observed in the
cross-domain case of \textbf{P2}.  

Second, while we notice that other cross-domain CF approaches, i.e., CBT and RMGM, also consistently outperform single-domain approaches, we can observe from Table~\ref{tab:performp1} that GTagCDCF performs better, i.e., by 3$\sim$4\% and
ca. 3.5$\sim$4.5\% over CBT on ML1 and LT, and by i.e., by 1.5$\sim$4.5\% and
ca. 3.5$\sim$7.5\% over RMGM on ML1 and LT. These results indicate that
exploiting explicit common knowledge between domains, such as social tags, could be more effective than relying on implicit common patterns between domains for the purpose of CDCF as used by the baseline approaches. This conclusion is also supported by the fact that our previous approach, TagCDCF, that is based on the same underlying rationale as GTagCDCF, also achieved substantial improvements over CBT and RMGM. In other words, it is the underlying paradigm of relying on social tags that makes the difference here, rather than the variant of the approach that uses this paradigm. 

Third, GTagCDCF significantly outperforms TagCDCF as
well. The relative improvement amounts to, for the case of \textbf{P1},
1.4$\sim$2.0\% on ML1 and 0.5$\sim$2.5\% on LT in terms of MAE, and for the case
of \textbf{P2}, 2.3$\sim$5.6\% on ML2 in terms of MAE and 6.8$\sim$14.7\% on LF
in terms of MAP. This result indicates that GTagCDCF succeeds in introducing the
frequency of user-tag relations and the frequency of item-tag relations for
deriving maximal benefit from common tags for improving recommendations in
each domain. This results, in combination with the previous one, shows that GTagCDCF outperforms all state-of-the-art CDCF approaches, which provides answer to our second research question.

\begin{table}
\tbl{Comparison of recommendation performance between GTagCDCF and
the baseline approaches on datasets in \textbf{P1}. 
\label{tab:performp1}
}
{%
\begin{tabular}{|c|c c|c c|c c|}
\hline
 & \multicolumn{2}{c|}{UPL=5}
 & \multicolumn{2}{c|}{UPL=10}
 & \multicolumn{2}{c|}{UPL=15}\\
 & \textbf{ML1} (MAE) & \textbf{LT} (MAE) & \textbf{ML1} (MAE) &
 \textbf{LT} (MAE) & \textbf{ML1} (MAE) & \textbf{LT} (MAE) \\
 \hline
\textbf{UBCF} & 0.833$\pm$0.009 & 0.857$\pm$0.009 & 0.785$\pm$0.004 &
0.795$\pm$0.005 & 0.766$\pm$0.003 & 0.771$\pm$0.004 \\ 
\textbf{PMF} & 0.831$\pm$0.010 & 0.771$\pm$0.009 & 0.771$\pm$0.010 &
0.761$\pm$0.006 & 0.753$\pm$0.010 & 0.769$\pm$0.010 \\ 
\textbf{CBT} & 0.792$\pm$0.009 &
0.729$\pm$0.010 & 0.752$\pm$0.003 & 0.694$\pm$0.008 & 0.737$\pm$0.003 &
0.682$\pm$0.006 \\ 
\textbf{RMGM} & 0.780$\pm$0.010 & 0.745$\pm$0.010 & 0.756$\pm$0.006 &
0.720$\pm$0.004 & 0.720$\pm$0.004 & 0.681$\pm$0.005 \\ 
\textbf{TagCDCF} &
0.777$\pm$0.008 & 0.709$\pm$0.004 & 0.735$\pm$0.004 & 0.674$\pm$0.003 & 0.719$\pm$0.003 & 0.663$\pm$0.002 \\ 
\textbf{GTagCDCF} &
\textbf{0.761}$^*$$\pm$0.008 & \textbf{0.691}$^*$$\pm$0.004 &
\textbf{0.721}$^*$$\pm$0.004 & \textbf{0.667}$^*$$\pm$0.002 &
\textbf{0.709}$^*$$\pm$0.003 & \textbf{0.659}$^*$$\pm$0.003
\\
\hline
\end{tabular}}
\begin{tabnote}%
\Note{Note:}{``$^*$'' denotes a significant
improvement of GTagCDCF over all the other approaches, according
to Wilcoxon signed rank significance test with p$<$0.005.} 
\end{tabnote}%
\end{table}%


\begin{table}%
\tbl{Comparison of recommendation performance between GTagCDCF and
the baseline approaches on datasets in \textbf{P2}. 
\label{tab:performp2}
}
{%
\begin{tabular}{|c|c c|c c|c c|}
\hline
 & \multicolumn{2}{c|}{UPL=5}
 & \multicolumn{2}{c|}{UPL=10}
 & \multicolumn{2}{c|}{UPL=15}\\
 & \textbf{ML2} (MAE) & \textbf{LF} (MAP) & \textbf{ML2} (MAE) & \textbf{LF}
 (MAP) & \textbf{ML2} (MAE) & \textbf{LF} (MAP) \\
 \hline
\textbf{UBCF} & 0.803$\pm$0.010 & 0.134$\pm$0.008 & 0.761$\pm$0.004 &
0.152$\pm$0.011 & 0.742$\pm$0.004 & 0.156$\pm$0.009 \\ 
\textbf{PMF} & 0.766$\pm$0.009 &
0.151$\pm$0.009 & 0.714$\pm$0.003 & 0.162$\pm$0.007 & 0.668$\pm$0.004 &
0.168$\pm$0.012 \\ 
\textbf{TagCDCF} & 0.748$\pm$0.009 & 0.162$\pm$0.010 & 0.686$\pm$0.003 &
0.175$\pm$0.011 & 0.653$\pm$0.003 & 0.190$\pm$0.013 \\ 
\textbf{GTagCDCF} & \textbf{0.707}$^*$$\pm$0.007
& \textbf{0.176}$^*$$\pm$0.008 & \textbf{0.655}$^*$$\pm$0.002 &
\textbf{0.201}$^*$$\pm$0.008 & \textbf{0.638}$^*$$\pm$0.003 &
\textbf{0.203}$^*$$\pm$0.013
\\
\hline
\end{tabular}}
\begin{tabnote}%
\Note{Note:}{``$^*$'' denotes a significant
improvement of GTagCDCF over all the other approaches, according
to Wilcoxon signed rank significance test with p$<$0.005.} 
\end{tabnote}%
\end{table}%

Finally, we investigate the impact of the number of common tags on the capability of CDCF to transfer knowledge between domains and benefit from it for recommendation. This investigation is done by taking a closer look at the comparison between the CDCF
approaches with tags, i.e., TagCDCF and GTagCDCF, and the CDCF approaches
without tags, i.e., CBT and RMGM. We compare their performance for users with
different numbers of used common tags, as shown in Table~\ref{tab:usertag}. Note
that we conduct this experiment only on the test set users of the LT dataset under
the condition of UPL=15, since the number of tags that the users assigned in ML1
varies in a much narrower range than LT so that it can hardly provide a
convincing observation. As can be seen from the table, for the users who used less than 5
common tags, the performance of TagCDCF and GTagCDCF does not obviously differ from that of CBT or RMGM. However, users who used more than 5 common tags
are able to benefit from TagCDCF and GTagCDCF, leading to performance superior
to that of CBT and RMGM. This observation provides further evidence that the
common tags could be more effective than implicit patterns mined from user preference for the purpose of
transferring knowledge for CDCF. Furthermore, it can be concluded that users who are active in tagging could particularly
benefit from the proposed approach. Since in our experiment the users who
used less than 5 common tags were in minority, the threshold of
the used common tags for GTagCDCF to be beneficial may not be difficult to reach in
practice. This analysis provides our answer to the third research question
regarding the dependence between the number of shared tags and the CDCF performance, but also to the fourth one, regarding the influence of user's individual tagging behavior on the benefit drawn from CDCF.

\begin{table}
\tbl{Performance comparison for users with different number of used common tags
in LT of \textbf{P1}.
\label{tab:usertag}}
{%
\begin{tabular}{c c c c c}
\hline
\# common tags (\# users) & CBT & RMGM & TagCDCF & GTagCDCF\\
\hline
 $<$5 (20) & 0.736 & 0.716 & 0.728 & 0.712 \\
5$\sim$10 (62) & 0.683 & 0.681 & 0.644 & 0.642 \\
11$\sim$20 (113) & 0.671 & 0.678 & 0.659 & 0.652 \\
20$\sim$50 (193) & 0.707 & 0.702 & 0.693 & 0.684 \\
$>$50 (175) & 0.665 & 0.664 & 0.641 & 0.639 \\
\hline
\end{tabular}}
\end{table}%

\subsection{Performance Comparison in the Three-domain Case}
As mentioned in Section~\ref{sec:setup}, for validating the robustness of
GTagCDCF to larger datasets and use cases with more than two domains, we
evaluate GTagCDCF in the three-domain case. The tradeoff parameters are tuned
to, $\alpha=10$ and $\beta=1$, based on the performance measured on the
randomly selected fold, as described in Section~\ref{sec:setup}. Then, the
performance of GTagCDCF is measured via 4-fold cross-validation. Due to the
relatively large size of the involved datasets, some baselines used in the two-domain case, as in the previous subsection, are too computational expensive to be deployed, such
as UBCF and TagCDCF. In addition, the baselines CBT and RMGM in the previous
subsection are not designed for the use cases with more than two domains. For
these reasons, only two single-domain baselines are employed for performance
comparison. One is PMF, which is the same as described in the previous
subsection. The other one is SVD++~\cite{Koren2008}, which is also a
state-of-the-art CF approach renowned from the Netflix Prize Competition.
SVD++ was developed for the purpose of rating prediction. We applied it in
the experimental settings in which we predict ratings, namely, ML-all and
LT-all datasets.
The implementation of SVD++ in our experiments is based on the publicly
available software
MyMediaLite\footnote{http://www.mymedialite.net/}~\cite{Gantner2011}.
The dimensionality of the latent features in both baselines is set to 10, the same as for GTagCDCF.
The related parameters in each of the baselines are also tuned based on the performance measured on the randomly selected fold, as described in Section~\ref{sec:setup}. 

The results are shown in Table~\ref{tab:perform3d}, from which we can observe
that GTagCDCF still achieves significant improvement over PMF and SVD++ in each
individual dataset, i.e., ca. 2\% improvement on the ML-all and the LT-all
datasets in terms of MAE, and ca. 3\% improvement on the LF dataset. These results provide an answer to our last research question and indicate
that GTagCDCF could be considered sufficiently robust for the use cases with large datasets
and multiple domains.

\begin{table}%
\tbl{Comparison of recommendation performance between PMF, SVD++ and GTagCDCF on datasets in the three-domain case. 
\label{tab:perform3d}
}
{%
\begin{tabular}{|c|c c c|}
\hline
 & \textbf{ML-all} (MAE) & \textbf{LT-all} (MAE) & \textbf{LF} (MAP) \\
 \hline
\textbf{PMF} & 0.610$\pm0.007$ & 0.656$\pm0.002$ & 0.416$\pm$0.005 \\
\textbf{SVD++} & 0.604$\pm0.004$ & 0.652$\pm$0.001 & N.A. \\
\textbf{GTagCDCF} & 0.596$^*$$\pm$0.001 & 0.644$^*$$\pm$0.001 &
0.430$^*$$\pm$0.003
\\
\hline
\end{tabular}}
\begin{tabnote}%
\Note{Note:}{``$^*$'' denotes a significant
improvement of GTagCDCF over all the baselines, according
to Wilcoxon signed rank significance test with p$<$0.005.} 
\end{tabnote}%
\end{table}%

\section{Discussion}
In this paper, we investigate the usefulness of social tags for addressing the
cross-domain collaborative filtering problem. We deploy the GTagCDCF approach
for this purpose that we developed by expanding the CMF framework for CF to a
multi-domain case. Experimental evaluation of GTagCDCF leads to important
insights regarding the effectiveness of different types of knowledge drawn from
different domains for the purpose of enabling CDCF. Social tags represent
explicit feedback data that may be generated by the users in a given content
domain. In this sense, the tags shared across domains can be seen as explicit
links among these domains. Compared to related
work~\cite{Li2009,Li2009a,Pan2010,Pan2011}, in which implicit correlations are
mined from user preference data, our experiments indicate that relying on
explicit links could be not only more effective for improving CDCF, but could
also make CDCF applicable to a wider range of use cases, for instance where the
items or users in different domains are not related to each other or when
different rating mechanisms are used across domains. Furthermore, increased
benefit based on deploying social tags does not require availability of
extensive sets of shared tags, but can be reached already with usual tagging
behavior of the users as observed in the current recommendation scenarios.

Regarding the GTagCDCF approach itself, our experimental results acquired for
two cross-domain cases, a homogeneous one and a heterogeneous one, demonstrate
that GTagCDCF could substantially outperform several state-of-the-art
CDCF approaches in both cases. This improvement is consistent for
users with various rating profiles. We also show that GTagCDCF outperforms our
previously proposed TagCDCF approach~\cite{Shi2011a}. This can be explained by
its ability to directly exploit user-tag relations and item-tag relations, while
not relying on any particular similarity measurement to capture the relations
between users or items from different domains. Finally, we validate the
robustness of GTagCDCF via the evaluation of its performance in the three-domain
case.

In our future work, we will work on generalizing our approach towards
exploiting the overall context that links different domains together for
improving recommendation in each domain. Under the ``context'', we understand a
mixture of information sources that link the domains in different ways. The
challenge here is to maximize the benefit for CDCF contributed by multiple types
of cross-domain links. This may also require considering different
recommendation frameworks apart from CMF.


\bibliographystyle{acmsmall}
\bibliography{MyRef_TIIS_TagCDCF}

\received{December 2013}{xxx xxxx}{xxx xxxx}


\end{document}